\documentclass[11pt]{article}

\usepackage[margin=2.6cm]{geometry}
\usepackage[colorlinks=true,linkcolor=blue,urlcolor=blue,citecolor=blue]{hyperref}

\usepackage{authblk}

\usepackage{amsmath,amssymb}
\usepackage{graphicx}

\newcommand{\rT}{\rho_{T}}
\newcommand{\pT}{p_{T}}
\newcommand{\Ne}{N_{e}}

\begin{document}

	\title{\textbf{Torsional pseudo-inflation beyond Einstein--Cartan}}
	
	\author[1]{Fernando Izaurieta\thanks{
			\href{mailto:fernando.izaurieta@uss.cl}
			{\texttt{fernando.izaurieta@uss.cl}}}}
	
	\author[2 ]{Samuel Lepe\thanks{
			\href{mailto:samuel.lepe@pucv.cl}
			{\texttt{samuel.lepe@pucv.cl}}}}

	\author[1]{Cristian Quinzacara\thanks{
			\href{mailto:cristian.quinzacara@uss.cl}
			{\texttt{cristian.quinzacara@uss.cl}}}}

	\affil[1]{Departamento de Ciencias Exactas, Facultad de Ingeniería, Universidad San Sebastián, Concepción, Chile}
	\affil[2]{Instituto de F\'isica, Pontificia Universidad Cat\'olica de Valpara\'iso, Av.\ Brasil 2950, Valpara\'iso, Chile}

\date{August 2026}

\maketitle

\begin{abstract}
\noindent
Torsion entered early-universe cosmology twice: through the nonsingular Einstein--Cartan bounce, where the quantum spin of fermions halts the contraction, and through the proposal that the expansion following that bounce could take over the duties of cosmic inflation. We revisit the second idea in a constructive spirit. First we compute its budget: for a Weyssenhoff spin fluid with barotropic source, $w\in\left[0,1\right]$, the accelerated window that follows the bounce spans $\Ne=\ln\left[4/\left(1+3w\right)\right]/3\left(1-w\right)\leq\tfrac{1}{3}\ln4\simeq0.46$ e-folds, and positive spatial curvature only shrinks it. The mechanism is sound; its torsional ``fuel'' dilutes as $a^{-6}$, and the engine stops two orders of magnitude short of inflationary needs. We then ask what torsion would need in order to sustain a quasi-de Sitter phase, and give a two-branch answer in Riemann--Cartan cosmology with nonminimal couplings, where torsion is fed by the couplings rather than by spin. The vectorial branch reproduces Palatini inflation, observationally alive. On the axial branch we exhibit an exact de Sitter solution sustained by a torsion condensate: constant axial torsion $f_{0}$, a linear Gauss--Bonnet coupling, and a potential fixed to $V_{0}=6f_{0}^{2}+\tfrac{5}{2}v^{2}$ by the Friedmann constraint. The solution is an attractor of the homogeneous dynamics, with eigenvalue exactly $-3H$. The phase ends when the slope $U^{\prime}$, a decreasing function of the roll speed at fixed flat potential, drifts down to the floor $5/16V_{0}$ and the condensate switches itself off; consistency of the effective theory pins the whole episode near the Planck scale, $H\gtrsim0.3$ in reduced Planck units. We studied this branch at background level; its perturbation spectrum remains open.
\end{abstract}


\section{Introduction}
\label{sec:intro}

Torsion has been knocking on inflation's door for forty years. Gasperini noticed in 1986 that the spin of matter, fed into the geometry through Einstein--Cartan (EC) gravity, could drive a phase of accelerated expansion~\cite{Gasperini1986}. Pop\l awski sharpened the idea into its boldest form: the torsion of spacetime, sourced by fermions, halts the contraction of the universe, produces a nonsingular bounce, and launches an expansion that could take over the duties usually subcontracted to an inflaton~\cite{Poplawski2010,Poplawski2012}. The proposal did what good geometric proposals do: it asked geometry to do a job normally assigned to a new field, it made the question quantitative, and it forced the community to ask how much inflation one can actually buy with spin. Its author has kept refining it since, adding quantum particle production precisely where the minimal mechanism runs out of breath~\cite{Poplawski2016,Unger2019}, and noting himself, with a candor worth imitating, where the spin-fluid description falls short of a scale-invariant spectrum~\cite{Poplawski1201}.

This paper takes that program at its word and asks the next question: what would it take for torsion to \emph{sustain} a quasi-de Sitter phase, rather than merely ignite one? The answer requires bookkeeping that is worth setting up carefully, because the word ``torsion'' covers three different kinds of theory. In EC gravity proper, torsion is algebraic, sourced by minimally coupled fermions, and locked to them: it does not propagate, and it dies wherever spin dilutes~\cite{Hehl1976,Trautman1973}. In more general Riemann--Cartan theories, nonminimal couplings source torsion even where spin is absent, while torsion remains algebraic and non-propagating. And in dynamical torsion theories, the torsion carries degrees of freedom of its
own~\cite{SNY2008,TorC2026}, a family that today includes the pseudoscalaron inflation models in which the axial mode, made propagating, is itself the inflaton~\cite{Salvio2022,DiMarco2024,GialamasTamvakis2025,Karananas2025}. These three theory ``boxes'' behave differently, and much confusion evaporates once the names are kept apart.

The first box is where the original proposal lives, and Sec.~\ref{sec:budget} computes its budget: the accelerated window of the spin-fluid bounce closes after at most $\tfrac{1}{3}\ln4\simeq0.46$ e-folds, for any barotropic source between dust and stiff matter, and positive curvature only makes it shorter. This is not a defect of the idea; it is the kinematics of a spin fuel that dilutes as $a^{-6}$. The number simply tells us where to look next.

The second box is where we look. In the nonminimally coupled scalar-field torsion cosmology of Ref.~\cite{Cid2018}, developed by one of the authors with Cid, Le\'on, Medina and Narbona, torsion is fed by the couplings $N\left(\phi\right)$ and $U\left(\phi\right)$, with no spin required, and it comes in two algebraic branches: a vectorial mode $h$, and an axial mode $f$ that survives only when the Gauss--Bonnet coupling is switched on. The seed of this line of thought was planted by Toloza and Zanelli~\cite{TolozaZanelli2013}, who coupled a scalar to the Euler form in Riemann--Cartan geometry and found that torsion and acceleration follow from the coupling alone; we are watering that seed at the inflationary end of the cosmic history. In a companion paper we watered it at the other end, showing that the same coupling-sourced torsion can impersonate the entire late dark sector \cite{companionA}, after a no-go for the diluting mode closed the alternative \cite{nogo}. The present paper completes the trilogy.

Two adjacent programs must be acknowledged at the outset to avoid confusion. The Einstein--Cartan inflation corpus of Shaposhnikov and collaborators~\cite{SSTZ2021,JHEP2020,MattersMatter,ShapoReview}, in parallel with L\r{a}ngvik et al.~\cite{LORR2021}, integrates out the algebraic torsion generated by nonminimal couplings linear in the curvature (with Holst and Nieh--Yan terms) and obtains the metric-to-Palatini family of Higgs inflation; our vectorial branch is, as we show in Sec.~\ref{sec:vectorial}, the minisuperspace shadow of that statement, and we import its observables~\cite{BauerDemir2008,RasanenWahlman2017,Tenkanen2020}. What that corpus does not contain is a scalar coupled to the Gauss--Bonnet invariant of the torsionful connection; there the axial mode survives as an algebraic, non-propagating condensate, and Sec.~\ref{sec:axial} shows it can hold an exact de Sitter phase up, with an exit built into the constraint structure. The pseudoscalaron models, for their part, promote the axial mode to a propagating field; ours never propagates, which keeps the theory lighter by one degree of freedom and keeps the exit algebraic. Neither neighbourhood, to our knowledge, occupies the corner we develop here.

A word on the title. We call the phase \emph{pseudo}-inflation because nothing in it slow-rolls in the usual sense: the scalar moves at constant speed, the potential is flat, and the near-de Sitter expansion is held by a torsion condensate that the coupling pays for. Whether the perturbations it seeds are viable is an open question; more on this in Sec.~\ref{sec:discussion}.

Conventions follow the companion paper \cite{companionA}: $\kappa_{4}\equiv8\pi G=1$, $c=1$, signature $\left(-,+,+,+\right)$, dots are cosmic-time derivatives, and primes are derivatives with respect to $\phi$. The deceleration parameter is given by $q=-1-\dot{H}/H^2$ and the first slow-roll parameter corresponds to $\epsilon = -\dot{H}/H^2$.

\section{The minimal mechanism and its budget}
\label{sec:budget}

\begin{figure}[t]
\centering
\includegraphics[width=0.9\textwidth]{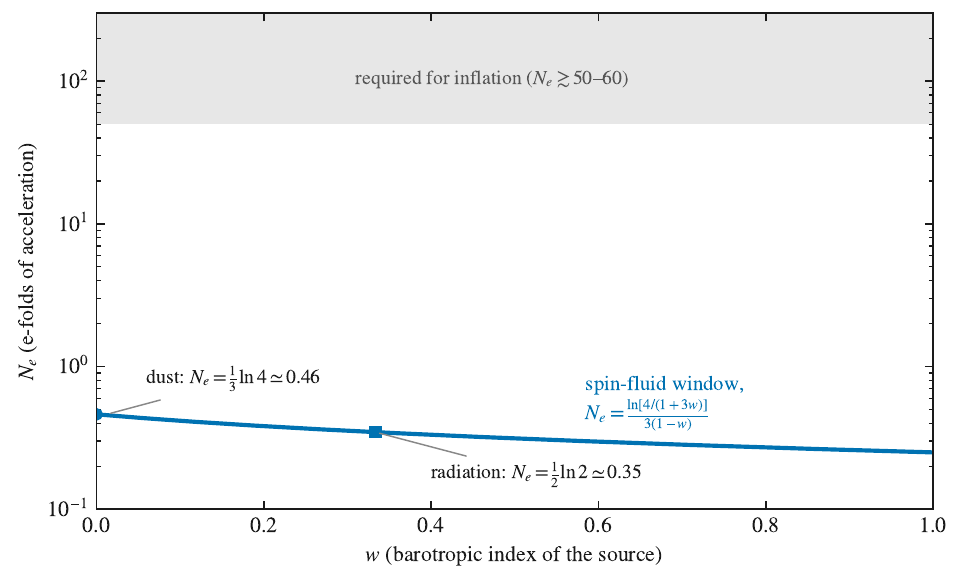}
\caption{The budget of the minimal mechanism. Accelerated e-folds of the spin-fluid bounce, Eq.~\eqref{eq:budget}, as a function of the barotropic index of the source, against the inflationary requirement. The window is kinematic: it depends on no parameter of the theory, only on the $a^{-6}$ dilution of the spin density. Positive curvature shrinks it further.}
\label{fig:budget}
\end{figure}

The Einstein--Cartan bounce is spin-fluid kinematics. For a Weyssenhoff fluid of unpolarized fermions~\cite{Weyssenhoff1947,NurgalievPonomariev1983,ObukhovKorotky1987,Brechet2008}, the macroscopic square of the spin density scales as $s^{2}\propto n^{2} \propto a^{-6}$, and the flat Friedmann pair reads
\begin{equation}
3H^{2}=\rho-\frac{s^{2}}{4}, \qquad 2\dot{H}+3H^{2}=-\left(p-\frac{s^{2}}{4}\right), \label{eq:spinfluid}
\end{equation}
with $p=w\rho$ and both sectors conserved. Two lines of algebra give the acceleration condition,
\begin{equation}
\ddot{a}>0
\quad\Longleftrightarrow\quad s^{2}>\left(1+3w\right)\rho , \label{eq:acccond}
\end{equation}
independent of spatial curvature, while the bounce sits at $s^{2}=4\rho$ (flat case; positive curvature moves it to $s^{2}<4\rho$). The accelerated window is therefore the interval in which $s^{2}/\rho$ falls from $4$ to $1+3w$, and since $s^{2}/\rho\propto a^{3w-3}$, its width is fixed by kinematics alone:
\begin{equation}
\Ne=\frac{\ln\left[4/\left(1+3w\right)\right]}{3\left(1-w\right)} \leq\frac{\ln4}{3}\simeq0.46 \quad\text{for }w\in\left[0,1\right]. \label{eq:budget}
\end{equation}
Dust gives $0.46$; radiation, the relevant case, gives $\tfrac{1}{2}\ln2\simeq0.35$; the stiff limit gives $1/4$. Positive spatial curvature narrows the window further.\footnote{Open curvature widens it instead: with $k=-1$ the bounce moves up to $s^{2}=4\rho+12/a^{2}$, so the window opens higher on the $s^{2}/\rho$ ladder. The gain is logarithmic and cannot rescue the budget (approaching $\Ne\sim50$ this way would demand a bounce dominated by curvature by a factor of order $e^{100}$, i.e., a nearly empty universe coasting rather than inflating), and the nonsingular bounces of the literature are closed~\cite{Unger2019}. We quote the closed case throughout.} Figure~\ref{fig:budget} puts the number where it belongs, two orders of magnitude below the $\Ne\gtrsim50$--$60$ that the horizon and flatness problems demand of an inflationary episode.\footnote{The precise requirement depends on the reheating scale; $50$--$60$ is the standard range~\cite{PlanckX}. Against a deficit of two orders of magnitude, the third significant figure is not the issue.}

Let us be precise about what this number does and does not say. It does not say the bounce fails: the bounce is the mechanism's one spectacular success, and it stands~\cite{Poplawski2012,Trautman1973,Stingley2026,Huang2025}. It does not say there is no gravitational repulsion: condition~\eqref{eq:acccond} is satisfied in a nonempty window, and near the bounce with $w\simeq-1$ sources the expansion is in fact super-accelerated, $q<-1$ (Appendix~\ref{app:landmarks}). What it says is that the repulsion is a transient. The spin density is a fuel that dilutes faster than anything else in the inventory, so the engine, however sound, carries a small fuel tank. It is direct to verify, along the same lines, that the slow-roll landmarks are displaced: with $s^{2}\neq0$ the limits $\epsilon\rightarrow0$ and $\epsilon\rightarrow1$ of the first slow-roll parameter $\epsilon=1+q$ no longer mark the de Sitter and Milne milestones of the standard story (Appendix~\ref{app:landmarks}), so the window cannot be stretched by tuning the source. And the $a^{-6}$ scaling is not an accident to be engineered away: it is equivalent to the adiabaticity of the fluid, as the continuity equation of~\eqref{eq:spinfluid} shows in one line.

On perturbations we have nothing to add to what the program's own author established: macroscopically averaged spin fluids do not hand over a scale-invariant spectrum~\cite{Poplawski1201}, the Dirac-field variant behaves differently at the level of the contact interaction~\cite{Kerlick1975,MagueijoZK2013}, and the Fermi-bounce line has shown what additional structure that route requires~\cite{Alexander2014,Alexander2015}. The averaging itself deserves its footnote.\footnote{Strict homogeneity and isotropy constrain which torsion components may survive averaging~\cite{Tsamparlis1979}, and for Dirac fields the four-fermion contact term enters with the opposite sign to the Weyssenhoff fluid~\cite{Kerlick1975}. We take the spin-fluid description at face value here precisely because our conclusion does not depend on its fine print: no value of $w$ rescues the budget~\eqref{eq:budget}.}


\section{Torsion beyond Einstein--Cartan} \label{sec:arena}

The tank we are after must not dilute, which rules out any source tied to a conserved particle number. Couplings do not dilute. The context is the same as in the companion paper~\cite{companionA}: the nonminimally coupled scalar-field torsion cosmology of Ref.~\cite{Cid2018}, with action
\begin{multline}
	S=\int\mathrm{d}x^{4}\sqrt{\left| g \right|}\left[\frac{N\left(\phi\right)}{2\kappa_4}R\left(g,\Gamma\right)-\frac{1}{2}\,\partial^{\lambda}\phi\partial_{\lambda}\phi-V\left(\phi\right)\right.\\
	\left.+U\left(\phi\right)\left(R^{2}-4R^{\mu}{}_{\nu}R^{\nu}{}_{\mu}+R^{\mu\nu}{}_{\rho\sigma}R^{\rho\sigma}{}_{\mu\nu}\right)+\mathcal{L}_{\mathrm{M}} \right] ,
	\label{eq:action}
\end{multline}
on Riemann--Cartan geometry: metric and connection independent, no torsionless condition imposed, the quadratic Gauss--Bonnet invariant built from the full connection with its nonvanishing torsion\footnote{This is also what separates the axial branch of this paper from metric Gauss--Bonnet inflation \cite{KGD2015}: there the GB density is constructed from the metric, $U\left(\phi\right)$ gravitates only through its gradient, and the coupling modifies the roll of an ordinary inflaton. Here the invariant sources a torsion component with no metric analogue.}, and $\mathcal{L}_{\mathrm{M}}$ some matter Lagrangian. In a flat Friedmann--Lema\^itre--Robertson--Walker background the torsion tensor reduces to the vector and axial components~\cite{Cid2018},
\begin{equation}
	T_{\lambda\mu\nu}=\left(g_{\mu\lambda}g_{\nu\rho}-g_{\mu\rho}g_{\nu\lambda}\right)h^{\rho}-2\sqrt{\left|g\right|}\,\epsilon_{\lambda\mu\nu\rho}f^{\rho}, \label{eq:ansatz}
\end{equation}
with comoving components $h_{0}=h\left(t\right)$, $f_{0}=f\left(t\right)$, and the field equations fix both algebraically. With $Z\equiv H+h$, the curvature scalar and the Gauss--Bonnet\footnote{With torsion the order of the contractions in \eqref{eq:action} matters: the Ricci tensor is no longer symmetric and the curvature tensor loses its pair-exchange symmetry, so, e.g., $R_{\mu\nu}R^{\mu\nu}\neq R^{\mu}{}_{\nu}R^{\nu}{}_{\mu}$. The invariant $\bar{\mathcal{G}}$ quoted here is the reduction of \eqref{eq:action} with its contractions as written (those of the Euler density).} invariant of \eqref{eq:action} evaluate on \eqref{eq:ansatz} to
\begin{equation}
\bar{R}=6\left(\dot{Z}+HZ+Z^{2}-f^{2}\right),
\qquad
\bar{\mathcal{G}}=24\left[\left(\dot{Z}+HZ\right)\left(Z^{2}-f^{2}\right)
-2fZ\left(\dot{f}+Hf\right)\right],
\label{eq:invariants}
\end{equation}
and, for a barotropic fluid $\left(\rho,p\right)$ of vanishing hypermomentum (classical matter; the spin fluid of Sec.~\ref{sec:budget} has left the stage), the complete set of field equations of \eqref{eq:action} can be brought to the form
\begin{align}
3N\left(Z^{2}-f^{2}\right)&=\rho+\frac{\dot{\phi}^{2}}{2}+V,
\label{eq:genfried}\\
N\left(2\dot{Z}+2HZ+Z^{2}-f^{2}\right)&=-p-\frac{\dot{\phi}^{2}}{2}+V,
\label{eq:genacc}\\
\ddot{\phi}+3H\dot{\phi}+V^{\prime}&=\tfrac{1}{2}N^{\prime}\bar{R}+U^{\prime}\bar{\mathcal{G}},
\label{eq:scalareom}\\
\dot{\rho}+3H\left(\rho+p\right)&=0,
\label{eq:matcons}
\end{align}
together with the two torsion constraints, from the variations with respect to $h$ and $f$,
\begin{equation}
N\,h=\dot{\phi}\left[4U^{\prime}\left(Z^{2}-f^{2}\right)
+\tfrac{1}{2}N^{\prime}\right],
\qquad
f\left(N-8U^{\prime}\dot{\phi}\,Z\right)=0 .
\label{eq:constraints}
\end{equation}
This is the complete system of Ref.~\cite{Cid2018}, regrouped: there, and in the companion paper \cite{companionA}, the metric pair is written instead as a pair of ordinary Friedmann equations fed by an effective torsion fluid $\left(\rT,\pT\right)$; the two forms agree term by term, with no constraint used in the regrouping.

Two remarks before the system goes to work. First, it is redundant in the familiar way: time-reparametrization invariance ties the display together, so once the constraints~\eqref{eq:constraints}, the scalar equation \eqref{eq:scalareom}, matter conservation \eqref{eq:matcons} and the Friedmann constraint \eqref{eq:genfried} hold, the acceleration equation \eqref{eq:genacc} is automatic. Second, the deformation of the standard cosmology is surgical. Equation~\eqref{eq:genfried} is the flat Friedmann equation under $3H^{2}\rightarrow3N\left(Z^{2}-f^{2}\right)$: the vectorial mode shifts the expansion rate that gravity responds to, the axial mode taxes it, and $N$ resets the effective Newton constant. The friction in \eqref{eq:scalareom} stays $3H\dot{\phi}$ rather than $3Z\dot{\phi}$ because the damping comes from the volume factor $a^{3}$, which knows the metric and not the connection. And setting $N=1$, $U=0$ forces $h=f=0$ in \eqref{eq:constraints}, collapsing the display to general relativity with a canonical scalar: the arena contains Einstein's cosmology as a corner, and every departure from that corner is written in $N^{\prime}$ and $U^{\prime}$.

The second constraint in \eqref{eq:constraints} is the fork on which everything in this paper turns. Either $f=0$, and torsion is a vectorial mode sourced by the coupling gradients, or
\begin{equation}
N=8U^{\prime}\dot{\phi}\,Z ,
\label{eq:branch}
\end{equation}
and the axial mode survives as a condensate sustained by the Gauss--Bonnet coupling. For $U=0$ the fork disappears and only the vectorial branch remains. Nothing here propagates: both modes are fixed pointwise by the fields, unlike the pseudoscalaron family; and both survive where spin is absent, so this is not Einstein--Cartan proper either.

\section{The vectorial branch: Palatini inflation in torsional clothing} \label{sec:vectorial}

Set $U=0$, drop the matter sector (we are in the early universe, and the scalar is the protagonist), and use the first constraint in \eqref{eq:constraints}, $h=\dot{N}/2N$. The generalized Friedmann equation \eqref{eq:genfried} then collapses, in one line, to
\begin{equation}
3N\,Z^{2}=\frac{\dot{\phi}^{2}}{2}+V,
\qquad
Z=H+\frac{\dot{N}}{2N}=\frac{d}{dt}\ln\left(a\sqrt{N}\right).
\label{eq:vectorial}
\end{equation}
The combination $a\sqrt{N}$ is the Einstein-frame scale factor of the conformal map $g\rightarrow Ng$, and in Einstein-frame time ($dt_{E}=\sqrt{N}\,dt$) Eq.~\eqref{eq:vectorial} becomes $3H_{E}^{2}=\tfrac{1}{2}\left(d\phi/dt_{E}\right)^{2}/N+V/N^{2}$: a canonical Friedmann equation with kinetic coefficient $1/N$ and potential $V/N^{2}$, and \emph{without} the $\tfrac{3}{2}\left(N^{\prime}/N\right)^{2}$ kinetic term that the metric formulation would add. That absent term is the entire difference between metric and Palatini inflation~\cite{BauerDemir2008,Tenkanen2020}, and its absence here is not an accident: the vectorial torsion is the conformal compensator that the Palatini formulation smuggles in through the connection, wearing, in Riemann--Cartan language, its honest name.\footnote{The statement that integrating out the algebraic torsion of nonminimal couplings linear in curvature lands on the Palatini side, and that Holst and Nieh--Yan terms interpolate toward the metric side, is due to Refs.~\cite{SSTZ2021,LORR2021}; Eq.~\eqref{eq:vectorial} is its minisuperspace shadow, and we display it because it can be read without integrating anything out.}

The consequence is pleasant: the vectorial branch is already known. It is Palatini inflation, a program with its own literature and its own observational record: quasi-de Sitter phases of arbitrary length, $n_{s}=1-2/N_{\star}$ for the attractor family (here $N_{\star}$ counts e-folds before the end of inflation, and has nothing to do with the coupling $N$), and a tensor-to-scalar ratio suppressed far below the metric-formulation value, comfortably inside current bounds~\cite{SSTZ2021,RasanenWahlman2017,PlanckX,BICEP2021}. The import is legitimate beyond the background level: the torsion is an auxiliary field, so eliminating it commutes with perturbing around any solution, and the fluctuations of this branch are Palatini fluctuations with no further calculation needed.

\section{The axial condensate}
\label{sec:axial}

Torsion sourced by couplings can not only sustain inflation; under the name Palatini, it quietly has been for years. What that literature leaves untouched is the other tine of the fork. So now let $U\neq0$ and take the branch \eqref{eq:branch}. To exhibit the mechanism in its purest form we choose the minimal hardware that supports it: constant curvature coupling, $N=1$, a \emph{linear} Gauss--Bonnet coupling, $U=u_{1}\phi$, and a flat potential, $V=V_{0}$. One function switched on, the other two switched off. The system then admits an exact de Sitter solution with every torsion ingredient constant,
\begin{equation}
a=e^{Ht},\qquad
\dot{\phi}=v,\qquad
f=f_{0},\qquad
h=h_{0},
\label{eq:condansatz}
\end{equation}
where the constraint pair \eqref{eq:constraints}--\eqref{eq:branch} and the field equations lock the constants together:
\begin{equation}
Z^{2}=v^{2}+3f_{0}^{2},
\qquad
h_{0}=\frac{Z^{2}-f_{0}^{2}}{2Z},
\qquad
H=\frac{Z^{2}+f_{0}^{2}}{2Z},
\qquad
u_{1}=\frac{1}{8vZ},
\qquad
V_{0}=6f_{0}^{2}+\frac{5}{2}v^{2}.
\label{eq:condensate}
\end{equation}
It is possible to verify that~\eqref{eq:condansatz}--\eqref{eq:condensate} solves all five equations of the system displayed in Sec.~\ref{sec:arena} (the torsion pair~\eqref{eq:constraints}, the scalar equation~\eqref{eq:scalareom} and the metric pair~\eqref{eq:genfried}--\eqref{eq:genacc}), and that the effective torsion fluid of Ref.~\cite{Cid2018} evaluates on it to $\rT=3H^{2}$, $\pT=-3H^{2}$: i.e. it is a cosmological ``constant'' assembled from dynamical parts. For $v=f_{0}=1$ the solution reads $Z=2$, $h_{0}=3/4$, $H=5/4$, $u_{1}=1/16$, $V_{0}=17/2$, in units $\kappa_{4}=1$.

It is worth pausing on what is doing what. The axial contribution enters the effective density with a positive sign, so $f_{0}^{2}$ is an energy reservoir; the branch condition~\eqref{eq:branch} says that reservoir exists only while the coupling flow $U^{\prime}\dot{\phi}$ holds it up; and the scalar, rolling at constant speed down no hill at all ($V^{\prime}=0$), is the clock that keeps the branch condition alive. Nothing slow-rolls; the expansion is held by geometry. This is why we insist on \emph{pseudo}-inflation: the phenomenology is inflationary, the mechanism is not the inflaton's.

It is fair to ask what the condensate buys that an ordinary inflaton on a flat potential would not. Take that inflaton: with $V$ exactly flat, the friction $3H\dot{\phi}$ kills the velocity as $a^{-3}$, the field freezes, and the universe settles into a de Sitter phase with no exit anywhere in its future. To end inflation one must bend the potential, and the bending must stay gentle enough to preserve slow roll; much of the craft of inflationary model building lives inside that compromise. Here the potential is never bent. The velocity survives because the Gauss--Bonnet source in~\eqref{eq:scalareom} balances the friction exactly, and the end comes from $U^{\prime}$ crossing a threshold, with $V^{\prime}=0$ throughout: the plateau of $U^{\prime}$ plays the role that the plateau of $V$ plays elsewhere. 

A satisfying but unintuitive exit is built in: promote $u_{1}$ to a slowly varying $U^{\prime}\left(\phi\right)$ and let the solution track it adiabatically (the attractor result below, Eq.~\eqref{eq:eigenvalue}, is what licenses this). The potential stays exactly flat, so $V_{0}$ is a fixed coupling of the theory and the drifting pair $\left(v,f_{0}\right)$ must respect it; solving the condensate relations \eqref{eq:condensate} at fixed $V_{0}$ gives
\begin{equation}
f_{0}^{2}=\frac{2V_{0}-5v^{2}}{12},
\qquad
Z^{2}=\frac{2V_{0}-v^{2}}{4},
\qquad
U^{\prime}=\frac{1}{4v\sqrt{2V_{0}-v^{2}}},
\label{eq:exit}
\end{equation}
with $U^{\prime}$ strictly decreasing in the roll speed on the whole physical branch. A large slope holds a slow, heavy condensate; as the scalar rolls into a region of declining $U^{\prime}$, the speed rises, the reservoir drains, and $f_{0}^{2}$ reaches zero at $v_{e}^{2}=2V_{0}/5$, where the slope meets its floor,
\begin{equation}
U^{\prime}_{\mathrm{floor}}=\frac{5}{16V_{0}}=\frac{1}{8v_{e}^{2}} .
\label{eq:floor}
\end{equation}
There the fork~\eqref{eq:constraints} snaps back to its vectorial tine: the condensate switches itself off, no tunneling and no cliff, and hands the universe to the $f=0$ branch, whose torsion then decays with the coupling flow.\footnote{At the snap the vectorial residue $h=4U^{\prime}\dot{\phi}Z^{2}$ survives and fades as the roll proceeds; the detailed handover, and with it the reheating history, needs the numerical treatment we defer to a future work. What the algebra already guarantees is that the endpoint is torsionless general relativity with a canonical scalar, and that the background crosses the snap smoothly: writing $U^{\prime}=\left(1+\delta\right)U^{\prime}_{\mathrm{floor}}$, the condensate melts linearly, $f_{0}^{2}\simeq\tfrac{4}{9}V_{0}\,\delta$, every background quantity continuous through $\delta=0$. What does break down near the snap is perturbation theory for the axial fluctuation, whose algebraic determination loses rank exactly there; the last stretch of the phase stands to fluctuations as the $\epsilon\rightarrow1$ endpoint of ordinary inflation does, and its physics belongs to the deferred reheating analysis, entropy budget included.} Inflation ends not because a potential steepens but because a constraint changes branch. The number of e-folds is set by the stretch of field over which $U^{\prime}$ stays above its floor, $\Ne=\int H\,\mathrm{d}\phi/v$: the tank is as large as the plateau of $U^{\prime}$ above the floor, and plateaus, unlike spin densities, do not dilute.

A worked profile makes the account concrete. Let us choose $V_{0}=17/2$, and let the slope descend as $U^{\prime}\left(\phi\right)=\tfrac{1}{16}\left[1+\left(\phi/\phi_{c}\right)^{2}\right]^{-1}$, so the phase starts at the benchmark condensate ($v=f_{0}=1$, $H=5/4$) and exits at $\phi_{e}=\sqrt{7/10}\,\phi_{c}$, where $U^{\prime}$ meets the floor $5/136$ with the field moving at $v_{e}=\sqrt{17/5}\simeq1.84$. Integrating $\Ne$ along the slaved branch gives $\Ne\simeq0.81\,\phi_{c}$: sixty e-folds cost $\phi_{c}\simeq75$, a field excursion $\Delta\phi\simeq62$ in Planck units, with the adiabaticity ratio $|U^{\prime\prime}\dot{\phi}/3HU^{\prime}|$ never exceeding $10^{-2}$ along the way. The excursion is large-field, as the operating point of Eq.~\eqref{eq:cutoff} below already announces.

The tracking above worked because the condensate attracts nearby histories. On the axial branch the constraints slave everything to the scalar velocity: Eq.~\eqref{eq:branch} fixes $Z=1/8u_{1}\dot{\phi}$, the first of Eqs.~\eqref{eq:constraints} fixes $h$, and Eq.~\eqref{eq:genfried} fixes $f^{2}$, so the homogeneous dynamics collapses to a single flow $\dot{p}=F\left(p\right)$ for $p\equiv\dot{\phi}$. The condensate is its fixed point, $F\left(v\right)=0$, and the eigenvalue comes out independent of every parameter of the hardware:
\begin{equation}
F^{\prime}\left(v\right)=-3H .
\label{eq:eigenvalue}
\end{equation}
Homogeneous deviations die as $\mathrm{e}^{-3Ht}=a^{-3}$, for any $v$ and $f_{0}$.\footnote{Within the homogeneous sector, that is: inhomogeneous perturbations remain the open front, and Sec.~\ref{sec:discussion} keeps them there. Three fine points. The eigenvalue is computed at frozen $u_{1}$; tracking a slowly varying $U^{\prime}$ is then legitimate whenever $U^{\prime\prime}\dot{\phi}/U^{\prime}\ll3H$, which is what ``plateau'' means here. The rate does not soften as $f_{0}\rightarrow0$, so the approach to the exit suffers no critical slowing. And deviations in $p$ large enough to drive $f^{2}$ through zero are not an instability either: they trigger the branch snap, which is the exit of Eq.~\eqref{eq:exit} arriving early.}

Here it is possible to explore some strong-coupling estimates. The operator $U\left(\phi\right)$ times the Gauss--Bonnet invariant carries a naive cutoff $\Lambda=1/U^{\prime}$, and the branch condition fixes $U^{\prime}=1/8vZ$, so $\Lambda=8vZ$ and, on the condensate,
\begin{equation}
\frac{H}{\Lambda} =\frac{v^{2}+4f_{0}^{2}}{16\,v\left(v^{2}+3f_{0}^{2}\right)} \in\left[\frac{1}{16v},\frac{1}{12v}\right]
\qquad\left(\kappa_{4}=1\right), \label{eq:cutoff}
\end{equation}
for any $f_{0}$. Keeping $H$ below $\Lambda$ at all therefore requires $v\gtrsim10^{-1}$ in Planck units, and a comfortable decade of hierarchy requires $v$ of order one; the benchmark $v=f_{0}=1$ sits at $H/\Lambda=5/64$. The squeeze deserves its numbers. Demanding $H/\Lambda\leq1/10$ forces $v\geq5/8$, and since $H\geq v/2$ on the branch (with equality at the exit, where $f_{0}=0$), the Hubble rate during the condensate phase is pinned above roughly $0.3$ in reduced Planck units: the corridor between ``inside the effective theory'' and ``comfortably below the Planck scale'' is narrow, and generic higher-curvature operators (which nothing here forbids!) are not obviously suppressed inside it. The problem with this is that a condensate rolling far below the Planck scale is outside its own effective theory. This phase operates near the top of the energy ladder or it does not operate, and we record that as a sharp feature of the mechanism rather than hide it; whether a high-scale condensate is compatible with the tensor bounds, and whether this narrow corridor survives the ultraviolet, are part of the yet unsolved perturbation problem.

To be explicit on this issue: background exactness is established; perturbations are not. Nothing in this work is a statement about the perturbations of the axial branch, neither their spectrum nor their stability. On this branch there is at least a clear and laborious route: integrate out the algebraic torsion, obtain the equivalent metric scalar-tensor theory (the Gauss--Bonnet sector will generate Horndeski-type operators), and compute the spectrum and the stability conditions with standard technology. It may be possible to proceed in a similar but a bit less laborious way using the operators introduced in Ref.~\cite{Barrientos2019b} in the context of Riemann-Cartan geometries. In any case, what we have proven here is that the condensate holds the background, as we discuss in Sec.~\ref{sec:discussion}.

\section{Discussion} \label{sec:discussion}

What survives of the original torsional-inflation vision is, we think, the best part of it. The bounce survives: nothing here touches the singularity avoidance that made Einstein--Cartan cosmology famous~\cite{Poplawski2012,Unger2019,Trautman1973}. The repulsion survives, window and all. What does not survive is the hope that a diluting spin density could also write the sixty e-folds; the budget \eqref{eq:budget} is two orders of magnitude short, for any barotropic source, and no tuning moves it. The constructive reading, which is the one this paper bets on, is that the proposal was pointing past its own minimal implementation: torsion can hold a quasi-de Sitter phase up, provided its source is the couplings, which do not dilute. On the vectorial branch this statement is not even new, it is Palatini inflation seen from the torsional side~\cite{SSTZ2021,LORR2021}, with its observables intact. On the axial branch it is new: an algebraic, non-propagating torsion condensate that holds an exact de Sitter background, Eq.~\eqref{eq:condensate}, attracts its homogeneous neighbourhood at the universal rate \eqref{eq:eigenvalue}, and abdicates on its own schedule, Eq.~\eqref{eq:exit}.

This leaves three open problems: first and largest, perturbations on the axial branch: the torsion is algebraic, so it can be integrated out exactly, and the Gauss--Bonnet sector will hand back a metric scalar-tensor theory with Horndeski-type operators whose spectrum and stability conditions could be computed with standard technology; until then the condensate is a background statement. Nothing about the vectorial branch waits on this: its spectrum is Palatini's, already computed and already viable~\cite{RasanenWahlman2017,Tenkanen2020}. One perturbative question we allow ourselves to flag as an expectation: the condensate background is parity-odd, $f_{0}\neq0$, so nothing obliges the two graviton helicities to propagate identically through it. At late times the same axial mode, fed there by fermion spin, rotates gravitational-wave polarization~\cite{Faraday2025}; whether its primordial cousin imprints a net chirality on the tensor spectrum (TB and EB correlations in the microwave background) is exactly the kind of question the deferred calculation exists to answer, and it is where this model would part company with most of its scalar-field impersonations. Second, the handover: the snap at the floor \eqref{eq:floor} is algebraic, but the subsequent decay of the vectorial residue, the draining of the condensate energy $3H^{2}$ into the scalar sector, and the reheating history it implies need numerical treatment; the pseudoscalaron literature has set the standard here, reheating included~\cite{Salvio2022,DiMarco2024}, and we intend to meet it in future work. Third, the health of the axial sector at the quantum level: parity, the coupling to fermion spin (which is also its opportunity: the same mode is the one that fermionic dark matter feeds at late times~\cite{Faraday2025,ILV2020,IL2020,CIL2020,Elizalde2023}), the ultraviolet fate of the near-Planckian operating point that Eq.~\eqref{eq:cutoff} imposes, and the radiative stability of the exactly flat potential, which loop corrections involving $U\left(\phi\right)$ have no reason to respect. On the classical side, this family of theories keeps gravitational waves luminal~\cite{Barrientos2019}, so no standard-siren objection applies.

One demarcation deserves repeating, because the theoretical neighbourhood is active. The pseudoscalaron models~\cite{Salvio2022,DiMarco2024,GialamasTamvakis2025,Karananas2025} make the axial mode propagate and appoint it inflaton; the Einstein--Cartan Higgs program~\cite{SSTZ2021,JHEP2020,MattersMatter,ShapoReview,LORR2021} scans couplings linear in curvature plus topological terms (and its scan does not include the Gauss--Bonnet invariant of the torsionful connection); and designer torsion histories imposed by hand~\cite{Guimaraes2021} have no Lagrangian origin. The corner occupied here, a non-propagating axial condensate sourced by $U\left(\phi\right)$, with an exit written into the constraint structure, is to our knowledge a new result. Its nearest ancestor is the scalar-Euler coupling of Toloza and Zanelli \cite{TolozaZanelli2013}; this work takes that original idea further.

This paper also closes a torsional trilogy on the phenomenology of Einstein-Cartan theories and nonminimal couplings. A no-go established that torsion which merely dilutes has no late-time story \cite{nogo}; a companion paper showed that torsion sourced by nonminimal couplings can impersonate the entire late dark sector, from cold dark matter to apparent phantom energy~\cite{companionA}; and here the same couplings, at the other end of the expansion history, hold up the kind of quasi-de Sitter phase that the spin-fluid mechanism ignites but cannot sustain.

\subsection*{Acknowledgments}
F.~Izaurieta acknowledges support from ANID-FONDECYT Regular grant 1262414, S.~Lepe from ANID-FONDECYT grant 1250969, and C.~Quinzacara from ANID-FONDECYT grant 11231238, all from the Government of Chile.

\subsection*{Data availability}
No observational data were created.

\appendix
\setcounter{equation}{0}
\renewcommand{\theequation}{A.\arabic{equation}}

\section{Displaced landmarks of the spin-fluid kinematics}
\label{app:landmarks}

For completeness we record the kinematic statements quoted in Sec.~\ref{sec:budget}, now with spatial curvature. With $k=+1$ the spin-fluid system gives
\begin{equation}
q=\frac{1}{2}\left[1+3\frac{w\rho-s^{2}/4+1/a^{2}}{\rho-s^{2}/4-3/a^{2}}\right],
\label{eq:appq}
\end{equation}
and it is direct to verify that the band $-1\leq q\leq0$ is equivalent to
\begin{equation}
\left(1+3w\right)\rho \leq s^{2} \leq 2\left[\left(1+w\right)\rho-\frac{2}{a^{2}}\right].
\label{eq:appband}
\end{equation}
The lower bound reproduces the curvature-independent acceleration condition~\eqref{eq:acccond}. The upper bound carries the surprise: for $w\rightarrow-1$ it becomes $s^{2}\leq-4/a^{2}$, which no real spin density satisfies. A spin fluid coexisting with a vacuum-like source therefore cannot sit \emph{inside} the band at all: near the bounce it is pushed below it, $q<-1$, into super-acceleration. There is more repulsion there, not less; what there is not is a quasi-de Sitter plateau.

The slow-roll landmarks shift in the same way. With $\epsilon=1+q$ one finds
\begin{equation}
\epsilon\rightarrow0:\left(1+w\right)\rho\rightarrow\frac{s^{2}}{2}+\frac{2}{a^{2}},
\qquad\qquad
\epsilon\rightarrow1:\left(w+\tfrac{1}{3}\right)\rho\rightarrow\frac{s^{2}}{3},
\label{eq:applandmarks}
\end{equation}
so neither limit marks the de Sitter or Milne milestone of the standard story while $s^{2}$ is alive; both milestones drift back to their textbook positions only as $s^{2}\propto a^{-6}$ dies. The landmarks, like the window, are sourced by the spin density. This is the kinematic content behind the budget of Fig.~\ref{fig:budget}, and the reason no choice of source stretches it.

\end{document}